\documentstyle[12pt]{article}
\makeatletter                                                           

\def\section{\@startsection {section}{1}{\z@}{-1.5ex plus -.5ex         
minus -.2ex}{1ex plus .2ex}{\large\bf}}                                 

\def\@thmcountersep{}                                                   

\long\def\@makecaption#1#2{\vskip 10pt \setbox\@tempboxa\hbox{#1. #2}   
   \ifdim \wd\@tempboxa >\hsize   
       #1. #2\par                 
     \else                        
       \hbox to\hsize{\hfil\box\@tempboxa\hfil}                         
   \fi}                                                                 

\def\runninghead{BRILL AND PIRK :\quad PICTORIAL HISTORY OF INSTANTONS}

\def\ps@headings{                                                       
 \def\@oddhead{\footnotesize\rm\hfill\runninghead\hfill}               
 \def\@evenhead{\@oddhead}                                              
 \def\@oddfoot{\rm\hfill\thepage\hfill}\def\@evenfoot{\@oddfoot} }      

\makeatother                                                            
\date{} 
\begin{document}
\pagestyle{headings}
\flushbottom
\title{0A Pictorial History of some Gravitational Instantons
\thanks {Research supported in part by the
National Science Foundation.}}
\author{ {\em Dieter Brill and
Kay-Thomas Pirk}\thanks{Department of Physics, University of Maryland,
College Park, MD 20742, USA and Max Planck Institut f\"ur Astrophysik,
8046 Garching, FRG.  The second author was supported by a Feodor-Lynen
fellowship of the Alexander von Humboldt-Foundation.}}
\maketitle
\vspace{-10pt} 

\begin{abstract}
  Four-dimensional Euclidean spaces that solve Einstein's equations
  are interpreted as WKB approximations to wavefunctionals of quantum
  geometry. These spaces are represented graphically by suppressing
  inessential dimensions and drawing the resulting figures in
  perspective representation of three-dimensional space, some of them
  stereoscopically. The figures are also related to the physical
  interpretation of the corresponding quantum processes.
\end{abstract}
\section{Introduction} 
Understanding General Relativity means to a large extent coming to
terms with its most important ingredient, geometry. Among his many
contributions, Charlie has given us new variations of this theme [1],
fascinating because geometry is so familiar on two-dimensional
surfaces, but so remote from intuition on higher-dimensional
spacetimes. The richness he uncovered is shown nowhere better than in
the 137 figures of his masterful text [2].

Today quantum gravity [3] leads to new geometrical features. One of
these is a new role for Riemannian (rather than Lorentzian) solutions
of the Einstein field equations: such ``instantons'' can describe in
WKB approximation the tunneling transitions that are classically
forbidden, for example because they correspond to a change in the
space's topology. In order to gain a pictorial understanding of these
spaces we can try to represent the geometry as a whole with less
important dimensions suppressed; an alternative is to follow the ADM
method and show a history of the tunneling by slices of codimension
one.

We can readily go from equation to picture thanks to computer plotting
routines, from the simpler ones as incorporated in spreadsheet
programs [4] to the more powerful versions of {\it
Mathematica}.\footnote{{\it Mathematica} is a trademark of Wolfram
Research Inc., 100 Trade Center Drive, Champaign, Il 61820-7237, USA.
We also used the program {\it showstereo.m} available by e-mail from
mathsource@wri.com. We thank Peter H\"ubner (MPI-Astro) and
Dr.-Ing.~Werner Rupp (DASA) for help with computer resources} We
therefore decided it would be fun to see how well the computer can
draw pictures associated with tunneling and instantons.  In Section 2
we recall the idea behind these by an example of a two-dimensional
potential. In the following sections we present and interpret several
general relativity instantons, both sliced and unsliced.
\section{Tunneling in Several Dimensions}
The one-dimensional potential (Fig.~1a)
$$V(x) = x^2 -x^3\eqno(1)$$
is typical of the class to which tunneling arguments are often applied.
\begin{figure}[h]
  \vspace{4.6cm}
  \caption{(a) Plot of the potential $V(x)$ with a resonance at $E_{res} =
  0.06605$. (Here we have put $\hbar^2/m = .01$.)  (b) Virtual state
  wavefunction $\psi(x)$ in this potential.}
\end{figure}
The exponential ``barrier penetration'' coefficient, which governs the
probability of escape for a particle that is initially trapped near
$x=0$, is related to the ``resonance'' (virtual state) solution of the
time-independent Schr\"odinger equation in this potential (Fig.~1b).
The maximum of this wavefunction occurs in the trapping region near
the relative minimum of the potential, and the oscillating ``tail'' in
the exterior region is very small compared to this maximum. This
property characterizes the virtual state. If the potential barrier is
sufficiently high, virtual states occur at rather well-defined
energies.

A priori the virtual state has nothing directly to do with the
tunneling problem: it is stationary, real-valued (at $t=0$) and yields
no net flux into or out of the potential's trapping region. (These
same properties also make this wavefunction easy to draw!) To obtain a
wavefunction with nonzero flux we need to use neighboring energy
states. The properties of their wavefunctions change rapidly as the
energy is varied away from resonance: to first order in the variation
the exterior phase changes, and to second order the interior amplitude
decreases. Fig.~2 shows two nearby energy states whose exterior phase
differs by $\pi/2$.
\begin{figure}[bht]
  \vspace{5cm}
  \caption{Wavefunctions $\psi_1$ (dashed) and $\psi_2$ (dash-dotted)
  at energies $E_1 = 0.0663$ slightly above resonance, and at $E_2 =
  0.0658$ slightly below resonance, showing the rapid phase change.
  The combination $\psi_1 - i \psi_2$ has a net outgoing flux that is
  also shown (solid).}
\end{figure}
 From these we can build an outgoing wave.  Because the amplitude of
these waves is essentially the same as that of the virtual state, the
tail of the virtual state is a measure of the outgoing flux. Of course
this measure is rather crude, giving us only the main exponential
factor in the decay probability.  Just how long it takes for the wave
to leak out depends on how rapidly the properties of the wavefunction
change with energy; thus this ``prefactor'' to the exponential is not
determined by the virtual state alone, and is rather more difficult to
compute.

In one dimension there is only one way for the particle to leak out of
the potential in Fig.~1a (namely, toward positive $x$).  To find in
more detail ``how'' the particle gets out of the trapping region, and
the analogous issues in tunneling of fields, we must consider at least
two-dimensional potentials.  Usually one treats the rotationally
symmetric case: angular momentum is then conserved, and the problem
reduces to a one-dimensional effective potential motion for each
angular momentum. Because of the symmetry no direction of tunneling
away from the trapping region is preferred over any other. But when
the potential is not symmetric, some tunneling ``paths'' can be
considerably more probable than others.\footnote{An optical analogon
is the observation of ``frustrated total internal reflection''[5]. The
potential barrier can be provided by the air space between the
hypotenuse faces of two $90^{\circ}$ prisms. More light (and at larger
wavelengths) gets through where this barrier is narrowest.}

A simple two-dimensional potential, in which the classical and the
quantum mechanical problem can be solved because the corresponding
equations are separable, is given by
$$V(x,y) = x^2 + y^2 - x^3 - 0.8 y^3 = V_1(x)+V_2(y).  \eqno(2)$$
(Here the factor $0.8$ was chosen only for convenience of plotting.)\,
Fig.~3 shows a plot of this potential.
\begin{figure}[htb]
  \vspace{5.8cm}
  \caption{3D plot of the potential $V(x,y)$. This, and some of the
  subsequent figures, are stereoscopic pairs similar to those found in
  the text {\it Methods of Theoretical Physics} by Morse and Feshbach.
  We know no better instructions how to view these figures than those
  found in the preface of that text.  }
\end{figure}
It has a trapping region that is separated from the ``exterior''
region ($x \gg 0, y \gg 0$) by a barrier with two saddle points of
different heights. The virtual state in this potential is simply the
product,
$$\psi(x,y) = \psi_1(x) \psi_2(y)$$ where $\psi_i$ is the virtual
state in potential $V_i$. This two-dimensional wave function is
plotted in Fig.~4a.
\begin{figure}[p]
  \vspace{17cm}
  \caption{(a)
  3D log square plot of the virtual state wavefunction for the
  potential of Fig.~3. As in Fig.~2, the interference maxima and
  minima have little to do with the physical outgoing wave, in the
  case that this wavefunction is used to represent the decay of a
  virtual state.  (b) Gray-level plot of the square of the
  wavefunction, showing the ``most probable escape path'' as a locus
  of relatively high density crossing the barrier region along the
  $x$-axis.  }
\end{figure}
In Fig.~4b the square of the wave function is indicated by increasing
gray levels.  We note that there is a path, namely along the $x$-axis,
on which the wavefunction is generally (and particularly at the end of
the exponential decrease) larger than along other paths leading from
the trapping region to the exterior. This is the ``most probable
escape path'' [6] that may be said to describe ``how'' the particle
most likely tunnels through the barrier, in this sense: if the system
were prepared in the virtual state and a position measurement were
made in the barrier region, the particle would most likely be found
near the most probable escape path. (Of course, after the measurement
the particle would no longer be in the virtual state.)

The tunneling behavior of a virtual state is well approximated by a
WKB wavefunction, and the analogous problem in {\it classical}
mechanics [7].  For example, the most probable escape path is given by
the ``bounce'' solution of particle motion in {\it imaginary time} ---
or equivalently, real time motion in the upside-down potential.  The
reader is encouraged to consider Fig.~3a turned upside down (for the
less agile reader we have provided Fig.~5) and to imagine how a
particle released at the central maximum would move in the resulting
potential. Fig.~6 shows
\begin{figure}[b]
  \vspace{6cm}
  \caption{Fig.~3 turned upside down.
  }
\end{figure}
\begin{figure}[b]
  \vspace{10cm}
  \caption{The equipotentials of $-V(x,y)$ and some particle orbits
  in this potential.}
\end{figure}
some of the orbits in such a potential. The bounce solutions are those
that have a strict turning point, where the particle velocity
vanishes; thus they are the only ones that reach the exterior zero
equipotential, where energy conservation allows the particle that
started at rest in the center to stop tunneling and again ``become a
classical particle.''

Along the bounce path we can solve a one-dimensional time-independent
Schr\"o\-dinger equation to find the fall-off of the wavefunction, and
hence obtain the tail amplitude and the exponential factor in the
decay rate. To essentially the same approximation we can use the WKB
estimate of this factor via the classical ``Euclidean'' action $S_E$
of the bounce path (in the potential $-V$). Since this is to be
computed for a motion in imaginary (``Euclidean'') time, the action $S
= iS_E$ itself is imaginary, and the lowest order WKB wavefunction,
$e^{iS} = e^{-S_E}$ becomes a decreasing exponential. If we compute
the action for the complete bounce, from the origin to the escape
point and back to the origin, we obtain twice the action for the most
probable escape path, which when exponentiated gives the probability
(rather than the amplitude).

In fact, we can get a continuous picture of the particle's history if
we allow the time parameter to change between real and imaginary at
turning points. Fig.~7 shows such a picture of the nonrelativistic
penetration history of the barrier of Eq.~(1) (but with time plotted
upward, as usual in relativity).
\begin{figure}[b]
  \vspace{6.2cm}
  \caption{The semiclassical history of the decay from the virtual
  state of the 1D potential $V(x)$. The particle starts at the
  classical ground state at $x = 0$, but moves in {\it imaginary}
  time, $it$, starting at $it = - \infty$, $x =
  \cosh^{-2}(it/\protect\sqrt{2m})$.  (This motion does not correspond
  to any passage of real, physical time, but signals the probabilistic
  nature of the process.) At the bounce (or ``nucleation'') time $it =
  0$ the motion stops momentarily at $x = 1$, and then continues in
  real time, $x = \cos^{-2}(t/\protect\sqrt{2m})$.  In real time the
  particle seems to appear suddenly at $x = 1$, and then continues on
  a classical escape orbit.  }
\end{figure}
We note that the infinite imaginary time needed for the tunneling has
nothing to do with the real time needed for the physical process
(which is vanishingly short in this approximation).  The WKB
wavefunction is given by the action evaluated along this entire
history. The Euclidean part of the motion contributes an imaginary
part to the action, and the real-time motion contributes a real part.
Thus the WKB wavefunction exhibits exponential decrease in the
Euclidean region, and oscillation in the real time region.  (However,
this wavefunction does not have continuous derivatives at the turning
points; this problem is usually solved by finding ``transition
formulas'', but we neglect it in this lowest approximation.)

In general, as in our example, there may be more than one bounce and
probable escape path. Because the probability is an exponential, the
one with the least Euclidean action is generally overwhelmingly more
probable (however, also see [8]). Of course, in order to get a
non-zero decay probability, this least action must be finite (this is
part of the definition of a bounce or instanton solution). Of the
three bounces shown in Fig.~6, the diagonal one has the largest action
(most improbable). In fact, it is really a combination of the other
two bounces, rather than a different way for the particle to escape.
This can be seen from the behavior of nearby particle orbits: they
converge toward this bounce, showing that there is a second zero mode
in the second variation of the action (the first zero mode is given by
an infinitesimal time translation). Because the orbits intersect, the
second variations vanish at the turning points, so that there will be
two variations that lower the action. Only bounces with {\sl one}
negative mode should be counted as independent decay channels.

Another kind of Euclidean solution (``instanton'') is useful for
potentials in which the escape region is replaced by a second trapping
region whose minimum is degenerate with the first (Fig.~8a).
\begin{figure}[ht]
  \vspace{8.5cm}
  \caption{(a) The potential $V(x) = -x^2 + x^4$ has two degenerate
  minima, separated by a tunneling region. (b) The semiclassical
  history of the tunneling in this potential, $x =
  (1/\protect\sqrt{2})\tanh(it/\protect\sqrt{m})$.  The particle
  starts from one of the minima at $it = -\infty$, and reaches the
  other minimum at $it = \infty$, with the entrire motion taking place
  in imaginary time.  }
\end{figure}
If one starts out with a wavepacket concentrated in one of the
regions, it will at a later time be concentrated in the other region
and vice versa, so that in general it has a fluctuating behavior. The
two lowest energy eigenstates that most importantly contribute to this
behavior are approximated by a sum resp. a difference between two WKB
wavefunctions (of the type $e^{S_E}$ and $e^{-S_E}$). These
wavefunctions can again be evaluated by finding the action of a
Euclidean solution of the classical equations of motion. But this
instanton does not exhibit a bounce. Instead the particle takes an
infinite imaginary time to move away from the center of the first
region, {\sl and} an infinite time to reach the center of the second
region (Fig.~8b). A corresponding two-dimensional potential and its
lowest energy wavefunction is shown in Fig.~9a.
\begin{figure}[p]
  \vspace{18cm}
  \caption{(a) A two-dimensional potential, $V(x,y)=-x^2+x^4+y^2$
  with two degenerate minima. (b) The ground state wavefunction is
  relatively large along the most probable connecting path.  }
\end{figure}
Again the wavefuntion is relatively large on the instanton path,
making it the ``most probable connecting path'' (Fig.~9b).

Thus, in a multidimensional setting we may again say that the
instanton tells us ``how'' the particle gets from one region to the
other during a fluctuation.  The instanton action will again give us
the main exponential factor in the probability that a particle
initially present in one region will be found in the other. In
particular, the existence of an instanton with finite action indicates
that the fluctuation in question does take place.  For the details of
the fluctuation, such as its frequency, one would again need
information about more than one energy eigenstate.\footnote{In this
connection, fluctuation means a transition between two or several
states that are ``classically allowed'' and have the same energy. We
do not mean the kind of virtual fluctuation that produces, for
example, a virtual pair in the vacuum state. (One could however say
that the barrier makes the virtual fluctuation real by preventing an
immediate return to the initial state.)}

One assumes that an analogous semiclassical approximation is valid in
field theory. That is, to see whether a classically forbidden
transition is possible in the quantum theory one tries to find a
finite action solution of the Euclidean field equations that connects
the relevant classical initial and final states of the transition. If
the connection is by a bounce solution, in which the initial state is
reached only asymptotically, but the final state occurs at the
turn-around ``point'' (surface of imaginary-time symmetry, or
nucleation surface [9]), the transition is interpreted as a decay. If
the connection is by an instanton (without a bounce), in which both
the initial and final states are reached only asymptotically, the
transition is a fluctuation. In general relativity, Euclidean
solutions of the Einstein equations are Riemannian (rather than
Lorentzian) manifolds.
\section{Gravitational Bounces}
If the classical initial state for tunneling has symmetry, the WKB
tunneling state may or may not exhibit the same symmetry. If the
Euclidean equations of motion admit a tunneling solution with the same
symmetry, we expect this solution to have the lowest action.  (If the
symmetry is broken by the tunneling, then there is no unique instanton
of smallest action, and one should sum over all of them.)  For
example, a constant electric field is invariant under boosts in the
field direction, and the instanton describing pair production by this
field [10] has the same invariance. The electric field is also
invariant under translation, but the instanton is not. However, the
tunneling events related by translation are all equally probable.

In accordance with this expectation the ``bubbles'' of true vacuum
expanding into false vacuum, both of which vacua are Lorentz
invariant, do exhibit invariance under the homogeneous Lorentz group,
and the corresponding Euclidean solutions is invariant under the
rotation group $O(4)$.  A simple example of a similar gravitational
situation is provided by the ``tunneling from nothing'' into a
deSitter universe [11], shown in Fig.~10 (with two dimensions
suppressed).
\begin{figure}[htb]
  \leavevmode \vspace{5.6cm}
  \caption{The semiclassical history of deSitter space,
  in the same spirit as Fig. 7.}
\end{figure}
As in Fig.~7, in the lower part time is imaginary, and in the upper
part it is real.  Each part solves the equation $G_{\mu\nu}= \Lambda
g_{\mu\nu}$ on its appropriate, Riemannian resp. Lorentzian, manifold.
The nucleation surface is the equator; reflection of either part about
this surface would give a complete manifold of each type, either a
Riemannian bounce (a complete 4-sphere) or a complete Lorentzian
deSitter universe (that bounces at some {\it minimum} radius).

To obtain a history of the deSitter evolution we normally slice the
upper part of Fig.~10 by horizontal, spacelike surfaces. For a history
of the tunneling it is natural to slice the lower part similarly by
horizontal planes. This history indeed shows nothing as long as the
plane is below the ``south pole'' of the hemisphere; when the plane
touches the pole, the universe originates as a point, then expands
into increasing 3-spheres (represented in the figure by circles) until
it reaches the maximum radius of the virtual evolution, which is also
the minimum radius of the deSitter space, and the real evolution
starts.

We have tacitly assumed that Fig.~10 is to be rotated about two other
axes to get the 4-manifolds (4-sphere, deSitter space) that we really
intended.  But we can just as well imagine the product of Fig.~10 with
$S^2$, to generate the Nariai metric [12] by tunneling from nothing.
In this case the nucleation surface, represented by the same
horizontal plane in Fig.~10 as before, has the topology $S^2 \times
S^1$ of a wormhole universe.  The Euclidean part of the action turns
out to be larger (corresponding to smaller tunneling probability) than
for tunneling into the more symmetric deSitter space [9] (cf.~the
$y-$axis vs.~the $x-$axis bounce of Fig.~6).

Further examples of $O(4)$ symmetric bounces are associated with
vacuum decay. Since the ordinary 4-dimensional Minkowski space vacuum
is stable [13], we have to consider compactified higher-dimensional
cases. To be realistic the spacetime should be at least 5-dimensional,
but in our figures we will have to suppress at least two of these
dimensions: each point will represent a 2-sphere (with metric
$d\Omega^2$ if it is a unit 2-sphere).\footnote {Actually, points
related by symmetry about the vertical plane perpendicular to the
paper correspond to the same 2-sphere.}

Figure 11a shows the ``5-dimensional Schwarzschild instanton,''
$$ds^2 = \bigl(1-(R/r)^2\bigr)d\phi^2 + \bigl(1-(R/r)^2\bigr)^{-1}dr^2
+ r^2 d\Theta^2 + r^2 \sin^2\Theta d\Omega^2, \quad r \geq R
\eqno(3)$$ used by Witten [14] to discuss the decay of the
Kaluza-Klein vacuum, with $\phi$ the compactification direction.
\begin{figure}[htb]
  \vspace{13cm}
  \caption{ (a) The semiclassical history of the decay of a compactified
  ``false'' 5D vacuum geometry is represented by the 3D region bounded
  above by the surface shown.  Each point of this region represents a
  two-sphere, obtained by rotating in two additional dimensions about
  the center of the figure. The front and back parts of the surface
  are to identified in such a way that points along the middle curve
  $C$ are regular (non-conical) origins at $r=R$ in the $r, \phi$
  plane. (b) The actual geometry obtained from a slice of Fig.~11a by
  a plane perpendicular to the paper, with the identification carried
  out.  }
\end{figure}
(Since the picture is only qualitatively correct, it could also
represent the 4-dimensional Schwarzschild instanton, which is usually
taken to describe the thermal properties of a Schwarzschild black
hole, but which in this connection would describe the decay of the
4-dimensional vacuum, compactified in one direction [15].) In the
lower part of this figure we have plotted $r$ and $\Theta$ in a
vertical plane, and $\phi$ in the orthogonal horizontal direction.
Since the latter is to be identified with period $2\pi$, the front and
back boundaries of the Figure are to be identified.  That this is
possible without singularity is shown in Fig.~11b, where this
identification is performed on a horizontal slice of Fig. 11a,
resulting in a ``test tube'' that is smoothly closed on one end, which
represents the metrically correct embedding of the $r, \phi$ section
of this metric.\footnote{Since the test tube is topologically ${\sl
I\!\!R}^2$, the whole metric of Eq.~(3) is ${\sl I\!\!R}^2 \times
S^3$, but one should remember that the ${\sl I\!\!R}^2$ does not have
the standard infinity, but instead becomes cylindrical --- like the
test tube.} The upper part of the Figure shows the analytic
continuation to a Lorentzian manifold.

The surface $\Theta = \pi/2$ in the Witten instanton (3) (or any
rotation of it) is a nucleation surface; it represents the final state
of the tunneling. Since the instanton is asymptotically flat, the
initial state, which it reaches asymptotically, is the vacuum. Once we
have chosen a nucleation surface, the initial vacuum is most
appropriately represented by the hyperplane at large distance from the
origin that does not intersect the nucleation surface. One can then
fill in other sections between these that give a reasonable tunneling
history [16]. At first the topology of these sections is ${\sl
I\!\!R}^3 \times S^1$, like that of a spacelike surface of the
compactified vacuum. When the minimum $r$ on such a section reaches
$R$, the section represents the singular instance at which the
topology changes.  Later sections have the topology ${\sl I\!\!R}^2
\times S^2$, like that of the final state. In the Figure this looks
like two disconnected spaces moving apart, but they are of course
connected through the two dimensions that were suppressed$^4$, so that
in fact a spherical hole has appeared in the final state, and expands
to infinity in the subsequent Lorentzian development.

As a final example we show the creation of two magnetically charged
Wheeler wormhole mouths by the magnetic field in an initially Melvin
universe [17]. This remarkable topology-changing bounce has two axes
of symmetry: one is the line joining the wormholes, and the other
corresponds to invariance under boosts of the created pair.  We
suppress the former to reduce the Euclidean space dimensions to three,
and show only the latter as a symmetry of the three-dimensional space
in which the figure is drawn, namely the rotational symmetry of
Fig.~12a about the horizontal axis, part of which is labeled
``horizon.''
\begin{figure}[p]
  \vspace{15cm}
  \caption{ (a) The imaginary-time part of the history of Wheeler wormhole
  production.  Beyond the top (nucleation) surface the real-time
  history could be represented in 2+1 Minkowski space, with the
  rotational symmetry replaced by a boost symmetry. (b) The wormhole
  at the nucleation surface is the geometry of the top surface of part
  (a), with the identification of the two circles carried out. (These
  circles are the vertical, circular grid line that is partically
  hidden behind the wormhole's throat.)  The grid lines can be taken
  to be equipotentials and field lines of the wormhole's field; to
  them should be added those of the Melvin universe background to get
  the net equipotentials and fields.  }
\end{figure}
(Except for this symmetry the Figure is qualitative only.\footnote{For
another way of representing this geometry --- in which two dimensions
are suppressed --- see Banks et al.~[18].}) For ease of visualization
we show only the lower, Euclidean half of the history --- the space
within the rectangular box and outside the semispherical cavities.  As
in the other figures, the upper part should consist of the analytic
continuation of the lower half to a Lorentzian spacetime.  Points on
the inner surface of the two cavities are to be identified by
reflection about the mid-plane.  The top horizontal plane is the
nucleation surface. It contains the axis of symmetry, where the
rotational Killing vector vanishes. The part between the cavities is
the wormhole's (Euclidean) Killing horizon.  Its Lorentzian extension
will as usual consist of two lightlike horizon surfaces, on which the
boost Killing vector that is the continuation of the rotational
Killing vector is lightlike. (The part of the axis outside the
cavities, not drawn in the Figure, corresponds to the ``Rindler
horizon'' that separates the two created wormhole mouths.)

On each horizontal section of Fig.~12a imagine a vector field like
that representing a laminar fluid flow from left to right, avoiding
the cavities. This can be taken as a representation of the magnetic
field.\footnote{In a spacetime the electromagnetic field tensor should
in general be represented by a honeycomb structure [2], not by lines.
But once one has chosen a ``spacelike'' surface, the usual electric
and magnetic field lines of course make sense.} In the bottom surface
this field is approximately constant and represents the field lines of
the (asymptotic) Melvin universe's central region [19]; away from this
central region the Melvin magnetic flux actually falls off with
distance. At the nucleation surface some of this flux crosses the
wormhole horizon and so has become trapped in the Wheeler wormhole.
To show this in more detail we have drawn in Fig.~12b only the top,
nucleation surface, but with the two circular curves in it identified
as required. The wormhole's Killing horizon now occurs, as expected,
at the narrowest part (throat) of the wormhole.

It is now not difficult to imagine other horizontal slices of Fig.~12a
to obtain the rest of the Euclidean history of the wormhole formation.
The singular slice is the one that first touches the cavities, so that
only a pair of points are to be identified, rather than two circular
curves.  It is the first slice in which some flux has been trapped,
and this trapped flux is conserved in the subsequent development. This
flux is maximal in the sense that, after the wormhole mouths have
moved far enough apart in their Lorentzian development to be compared
to single black holes, they correspond to extremal
Reissner-Nordstr\"om black holes. It is interesting that the spacelike
distance through the wormhole is finite on the nucleation surface; it
increases as the wormhole mouths move apart after nucleation, and
asymptotically approaches the infinite value that one expects from the
``horn'' structure of the isolated Reissner-Nordstr\"om black hole
[20]. It seems that other interesting features of the Lorentzian
analytic continuation have been explored only to a limited extent, and
they are too far removed from the instanton to be treated here [18].

\section{Fluctuation Instantons}

The solution of the Euclidean Einstein-Maxwell equations,
$$ds^2 = V^{-2} dt^2 + V^2 (dx^2 + dy^2 + dz^2) \qquad V = \sum
m_i/r_i, \qquad *F = d(V^{-1}) \wedge dt \eqno(4)$$ can be interpreted
as an instanton [21]. Here $r_i,\ i=1...n$ are the distances in flat
Euclidean 3-space from $n$ different origins to the field point. This
solution is similar to the multi-extremal Reissner-Nordstr\"om
solution, except that it has no asymptotically flat region; instead it
becomes cylindrical in the limit $r \rightarrow \infty$, just as it
does in each ``horn'', $r_i \rightarrow 0$. If we place all the
origins in the $xy$ plane and suppress the $z-$ and $t-$directions,
the geometry of the remaining dimensions can be shown embedded in flat
Euclidean space as in Fig.~13.
\begin{figure}[htb]
  \vspace{6.5cm}
  \caption{Imaginary time history of a fluctuation by which a maximally
  charged black hole throat splits into two.  }
\end{figure}
(The Gaussian curvature of this 2-surface is negative, reaching zero
asymptotically; it appears remarkably difficult to find an accurate
and unique embedding of such a surface, so Fig.~13 is correct only in
its main features.) The way {\it Mathematica} sliced this figure
suggests that this is an instanton that interpolates between a single
universe and two or several daughter (or baby) universes.  Indeed, the
geometry and fields of Eq.~(4) approach the Euclidean version of a
Bertotti-Robinson universe [22] in all asymptotic regions. The fact
that there is no bounce or nucleation surface, and that all
connections to Lorentzian solutions are made in the asymptotic
regions, indicates that this is a fluctuation-type instanton.

Many of the details of this instanton have been given elsewhere [21],
where it is argued that it should also describe the quantum
fluctuations of the region near the horizon of an extremal
Reissner-Nordstr\"om black hole. Here we want to show the result of
slicing this instanton for $n=2$ and $m_1 = m_2$ by 3-surfaces. A
simple choice consists of the surfaces on which $V$ is constant. These
slices are shown in Fig.~14.
\begin{figure}[b]
  \vspace{.1cm}
  \caption{(next page)
  Slices of constant $V$ of the instanton of Fig.~13, with only one
  direction suppressed, (a) ``before'' the topology change (b)
  singular slice at the instant of topology change (c) ``after'' the
  topology change.}
\end{figure}
They are metrically accurate (except for the suppression of the axis
direction).  For small $V$, and hence large $r$, these surfaces of
constant $V$ have the topology of a single sphere. For large $V$, one
or the other of the $r_i$ has to be small, so one obtains two separate
spheres. The critical, singular surface occurs when $V = m/d$, where
$d$ is the distance in Euclidean 3-space between the two origins.
Depending on the interpretation, Fig.~14 then gives the imaginary-time
history of the break-up of a Bertotti-Robinson universe, or of an
extremal Reissner-Nordstr\"om black hole's horizon.
\section{Conclusions}

When instanton solutions were first investigated in general
relativity, they were regarded primarily as a mathematical device, not
to be interpreted in the same way as Lorentzian solutions --- it was a
bold step even to draw a Riemannian and a Lorentzian manifold
connected in the same diagram.  Since then it has become clear that
many of the traditional methods work in both arenas. In particular, we
have seen above that slicing by 3-dimensional surfaces can help the
physical interpretation, just as it did for classical relativity when
ADM pioneered this method.  \vfill \pagebreak \newpage ${^{}}$
\pagebreak

\bibliographystyle{plain}

\end{document}